\documentstyle[12pt]{article}
\def\al{\alpha}
\def\be{\beta}

\def\de{\delta}
\def\ep{\epsilon}

\def\la{\lambda}

\def\mn{{\mu\nu}}

\def\half{{\textstyle{1\over 2}}}
\def\frac#1#2{{\textstyle{{#1}\over {#2}}}}

\def\lsim{\mathrel{\rlap{\lower4pt\hbox{\hskip1pt$\sim$}}
    \raise1pt\hbox{$<$}}}
\def\gsim{\mathrel{\rlap{\lower4pt\hbox{\hskip1pt$\sim$}}
    \raise1pt\hbox{$>$}}}
\def\sqr#1#2{{\vcenter{\vbox{\hrule height.#2pt
         \hbox{\vrule width.#2pt height#1pt \kern#1pt
         \vrule width.#2pt}
         \hrule height.#2pt}}}}

\def\lrprtmu{\stackrel{\leftrightarrow}{\partial_\mu}}

\newcommand{\beq}{\begin{equation}}
\newcommand{\eeq}{\end{equation}}
\newcommand{\bea}{\begin{eqnarray}}
\newcommand{\eea}{\end{eqnarray}}

\renewenvironment{thebibliography}[1]
 { \rm
   \begin{list}{\arabic{enumi}.}
    {\usecounter{enumi} \setlength{\parsep}{0pt}
     \setlength{\itemsep}{3pt} \settowidth{\labelwidth}{#1.}
     \sloppy
    }}{\end{list}}
 
\begin{document}
\titlepage
 
\begin{flushright}
{NCF 5\\}
{Jan 2012\\}
\end{flushright}

\vglue 1cm
	    
\begin{center}
{{\bf Classical Lagrangians for Momentum Dependent Lorentz Violation\\}
\vglue 1.0cm
{Don Colladay and Patrick McDonald \\} 
\bigskip
{\it New College of Florida\\}
\medskip
{\it Sarasota, FL, 34243, U.S.A.\\}
 
\vglue 0.8cm
}
\vglue 0.3cm
 
\end{center}
 
{\rightskip=3pc\leftskip=3pc\noindent
Certain momentum-dependent terms in the fermion sector of the Lorentz-violating Standard Model Extension (SME) yield solvable classical lagrangians of a type not mentioned in the literature.  These cases yield new relatively simple examples of Finsler and pseudo-Finsler structures.
One of the cases involves antisymmetric $d$-type terms and yields a new example of 
a relatively simple covariant lagrangian.} 

\newpage
 
\baselineskip=20pt
 
{\bf \noindent I. INTRODUCTION}
\vglue 0.4cm
 
The Standard Model Extension (SME) is an effective
field theory that has been formulated to include all
possible Lorentz-violating background couplings to Standard Model
fields \cite{ck}.  The SME framework exhibits many of the properties
of standard quantum field theories including gauge invariance,
energy-momentum conservation, causality and stability (in concordant 
frames)\cite{kle}, observer Lorentz invariance and hermiticity. 
In addition, numerous renormalizability properties of the theory have
been established \cite{klp1,klp2,cm1}, etc, and various implications for
particle theory, gravity \cite{kosgrav} (Lorentz violation provides an alternative
means of generating theories of gravity \cite{kpott}) and cosmology
(see  \cite{kmewes} for a study of the relationship of Lorentz
violation to  cosmic microwave background data) have been
discussed. There are a number of SME limits which yield exactly
solvable dispersion relations \cite{cmm}.  Extensive calculations
using the SME have led to numerous experiments which have been
performed to bound the LV effects predicted by the theory.  These
experiments involve numerous aspects of the SM (bounds associated to
electrons, photons, neutrinos, and hadrons, etc).   An exhaustive list
and a brief discussion of relevant  experimental results are contained
in a well-maintained set of data  tables \cite{neil} (see also
\cite{cpt98}). 

A geometric framework in which the classical limit of the SME can be naturally formulated has
recently been proposed.  This framework, Finsler geometry, is a
natural extension of Riemannian geometry (see section 2 for a brief
introduction).  Effectively employed by 
Randers \cite{randers} to study dynamics of charged particles coupled to a
fixed background field, Finsler geometry has been used to study a
number of special limits of the SME where certain background fields
are allowed to couple to standard fields \cite{kosfins}, \cite{aknr}.  The
associated SME lagrangians lead to Finsler structures which do not, in
general, define Randers-type geometries.  

The recent announcement of the OPERA group \cite{opera}, if confirmed, point
 towards the necessity of reworking the fundamental assumptions
 underlying the construction of at least the neutrino sector of the Standard Model.  Of particular
 interest is the assumption of Lorentz invariance and the possible
 observable signatures which might be expected should Lorentz symmetry
 be broken.  Reconciling the modified dispersion relation with other observed
 physics has proven challenging due to apparent inconsistencies including
 radiative processes \cite{colglash} and effects on pion decay properties \cite{brett},
 and generic problems with single parameter modifications \cite{kosneut}.
 While it is unclear how to resolve these inconsistencies at present, it seems
 that some modification of the dispersion relation for neutrinos will be unavoidable 
 if the OPERA result is verified.  Attempts at evading these conventional problems
 with radiative decays using Finsler space have recently been proposed \cite{clw}. 
 
The purpose of this note is to investigate new limits of the SME in
the context of Finsler geometry.  Using a factorization of the
dispersion relation \cite{cmm}, we study limits of the SME which
involve momentum dependent couplings.  These limits yield classical
lagrangians and exactly solvable models, new to the literature, which
define associated pseudo-Finsler structures away from a singular set
(whose structure is of independent interest).  We show that Wick
rotation leads to Finsler geometries which are not Randers-type
geometries.  

The remainder of the paper is structured as follows.  In the next
section we provide an introduction to the required background material
and establish notation.  In the third section we derive classical
lagrangians for new limits of the SME and discuss the associated
velocity-momentum relationship. In the fourth section we define
Finsler structures naturally associated to our limits of the SME.  We conclude
with a discussion which includes directions for future work.

\vglue 0.6cm
{\bf \noindent II. NOTATION, CONVENTIONS AND BACKGROUND}
\vglue 0.4cm

The SME Lorentz-violating Lagrangian
for a single spin-$\frac{1}{2}$ fermion is given by
\beq
{\mathcal L} =  {i \over 2} \bar{\psi}\Gamma^\mu \lrprtmu \psi -
\bar{\psi}M \psi ,
\label{lag2.1}
\eeq
where 
\bea
\Gamma^\nu & = & \gamma^\nu + c^{\mu\nu}\gamma_\mu + d^{\mu\nu}\gamma_5 \gamma_\mu
+ e^\nu + if^{\nu}\gamma_5 + \frac12 g^{\lambda \mu \nu}
\sigma_{\lambda \mu}, \label{Gamma2.1} \\
M & = & m + a_\mu\gamma^\mu + b_\mu\gamma_5 \gamma^\mu + \frac12
H_{\mu\nu}\sigma^{\mu\nu}\label{mass2.1}.
\eea

The coefficients $c_{\mu\nu}$,  $d_{\mu \nu}$, $ e_\nu$,  $f_\nu$,
$g_{\lambda \mu \nu}$, $a_\mu$, $b_\mu$, and $H_{\mu\nu}$ governing
Lorentz violation are assumed small.  The requirement of hermiticity of the lagrangian forces
the parameters to be real.  In addition, the parameters 
$c_{\mn}$ and $d_{\mn}$ can be taken to be traceless, $H_{\mu\nu}$ 
  antisymmetric, and $g^{\lambda\mu\nu}$ antisymmetric in the first
  two indices.  The parameters $a_{\mu}$, $b_\mu$, and $H_{\mu\nu}$
  have the dimension of mass, while the remaining parameters are
  dimensionless. 
  
As mentioned above, the SME exhibits many of the properties of
standard quantum field theories including gauge invariance,
energy-momentum conservation, causality, stability, observer Lorentz
invariance, hermiticity and power counting renormalizability.  In
addition, any theory that generates the SM and exhibits spontaneous
Lorentz and CPT violation contains the SME as an appropriate limit
\cite{ck}.  

The Dirac equation associated to the Lagrangian (\ref{lag2.1}) is
given by  
\beq
  (i\Gamma^\nu \partial_\nu - M)\psi =0  ,
\eeq
or, in momentum space coordinates (using $\psi(x) = e^{- i p \cdot x} u(p)$ for now)
\beq
  (\Gamma^\nu p_\nu - M)\psi =0. \label{dirac2.1}
\eeq
The Dirac operator $(\Gamma^\nu p_\nu - M)$ is a $4\times 4$ matrix with
complex entries.  The dispersion relation characterizes the null space
of the Dirac operator and is given by  
\beq
  {\rm det} (\Gamma^\nu p_\nu - M) =0. \label{dispersion2.1}
\eeq

Expression (\ref{dispersion2.1}) describes the zeroes of a fourth
order polynomial in $p^0$ whose coefficients depend smoothly on the
Lorentz-violating parameters and on the momentum vector.  
While the explicit covariant form of this
dispersion relation is readily available \cite{kle}, the complexity of the
general expression impedes the quantitative analysis required to
produce meaningful physical predictions in the presence of Lorentz
violation.  One method for addressing this situation involves defining
special limits of the SME by constraining certain combinations of the
coefficients appearing in (\ref{Gamma2.1}) and (\ref{mass2.1}) to be
zero.  A well-studied example of such a special limit is the momentum
independent $ab$-limit of the SME in which all coefficients except $a$ and
$b$ in (\ref{Gamma2.1}) and (\ref{mass2.1}) are tuned to zero.
The Hamiltonian for this limit can be implicitly defined from the 
covariant dispersion relation
\beq
((p-a)^2- m^2 +b^2)^2 - 4 (b \cdot (p-a))^2 + 4 m^2 b^2 = 0 ,
\eeq
which lends itself to facile analysis.  Calculating the implicit derivative
$u^i  = -u_0 \frac{\partial p^0}{\partial p_i}$ and combining the 
resulting three equations and the dispersion relation 
into a single equation for $L =-  u \cdot p$ 
yields an octic polynomial in $L$ that factors and yields directly to
the remarkably simple classical particle lagrangian 
\cite{aknr}
\beq
L_{ab}  =  -m\sqrt{u^2}  - a \cdot v \mp \sqrt{(b\cdot u)^2 -
  b^2u^2}. \label{ablagrangian} 
\eeq
It is interesting to note that an alternative action for this theory
has recently been discovered that eliminates the need for the
square roots and implements lagrange multipliers \cite{romero},
but this will not be pursued in the present work.

In addition to providing for a detailed analysis of classical particle propagation 
properties, formula
(\ref{ablagrangian}) suggests a framework for the
classical theory within the context of Finsler geometry.  

Let $M$ be a $C^\infty$
manifold with tangent bundle $TM,$ a {\it Finsler structure for $M$}
is a function $F: TM \longrightarrow [0,\infty)$ satisfying 
\begin{enumerate}
\item  $F$ is $C^\infty$ away from the zero section of $TM.$
\item  $F(x,\lambda u) = \lambda F(x,u)$ for all $\lambda >0.$ 
\item  The Hessian 
\beq
g_{ij} = \left( {1 \over 2}F^2 \right)_{u_i u_j},
\eeq
where the subscripts indicate conventional
differentiation, is positive definite at every point of $TM\setminus 0.$
\end{enumerate}
As an example, to study the dynamics of relativistic electrons in a
background magnetic field, Randers \cite{randers} introduced the Finsler
structure  
\beq
F(x,u) = \sqrt{u^2} +A_i(x) u^i ,
\eeq
where $A$ is a magnetic vector potential.
In more generality, if $M$ is a $C^\infty$ manifold with
Riemannian structure $r_{ij}$ and $A$ is a 
one-form on $M,$ then 
\beq
F(x,u) = \sqrt{r_{ij}(x)u^iu^j} + A_i(x)u^i ,
\eeq
defines a Finsler structure on $M$ which is called Randers.  Randers 
structures are classified by the Matsumoto torsion, an invariant
constructed using derivatives of the Finsler structure (for an
introduction to Finsler geometry, see \cite{BCS} or \cite{ZS}).  

The formal similarity between electrons moving in a background field
and particles coupling to tensors which break Lorentz symmetry
suggests that Finsler spaces may provide a geometric framework for the
SME.  Indeed, a connection between the modified dispersion relations
arisings in the SME and Finsler geometry goes back at least to
Bogoslovsky \cite{Bo1}, \cite{Bo2}.  Kostelecky has developed these
ideas for the $ab$-limit of the SME.  More precisely, in \cite{kosfins}
the classical lagrangian (\ref{ablagrangian}) is used to construct a
Finsler structure which is investigated in detail.  Fixing a
Riemannian metric $r_{ij}$ on a background spacetime manifold $M,$ let
\beq
F_{ab}  =  \sqrt{y^2} + a\cdot y \pm \sqrt{ b^2y^2 - (b\cdot y)^2} ,
\label{ablagrangian2.2} 
\eeq
where the dot products are taken using $r_{ij}$, and $y$ is the
velocity vector in $TM$.  The Finsler function $F$ is therfore 
$C^\infty$ and positive on $TM\setminus S$ where $S$ is comprised of
the zero section as well any other point at which $F$ vanishes. 
It has become customary to refer to this space as "ab-space".
Tuning $a$ to zero produces a Finsler structure on $TM\setminus S$ which is
non-Randers.  
Other relatively simple spaces have been constructed similarly using $f$, $c$, and $e$, as
well as certain limits of $H$.

To carry out a similar investigation for other SME limits, we begin
with an investigation of the modified dispersion relation developed in
\cite{cmm}.  To combine parameters with similar behavior in the dispersion
relation, the following definitions are implemented
\begin{eqnarray}
{d}^i_1   =  d^{0i}  & \hspace{.25in}& 
{d}^i_p  =  d^{ij}p_j  \\ 
{H}^i   =  H^{0i}  & \hspace{.25in}& 
{G}^i  =  g^{0ij}p_j  \\ 
{h}^i   =  1/2 \epsilon^{ijk}H^{jk}  & \hspace{.25in}& 
{g}^i  =  1/2 \epsilon^{ijk}g^{jkl}p_l ,
\end{eqnarray}
and
\begin{eqnarray}
\alpha_0  =  b^0 + \vec {d}_1\cdot \vec {p}   & \hspace{.25in}& \vec {\alpha}  = 
\vec {H} - \vec {G} \label{alphadef3.1} \\
\vec \delta_1  =  \vec {b} +\vec {d}_p +(\vec {g} - \vec {h}) & \hspace{.25in} & 
\vec \delta_2  =  - \vec {b} - \vec {d}_p +
(\vec {g} - \vec {h}). \label{epsilondef3.1}  
\end{eqnarray} 

When $\vec {\delta_1} = - \vec {\delta_2}$ and $\alpha_0=0,$ the associated
dispersion relation becomes \cite{cmm}
\begin{equation}
p_0^2 =   \vec p^2 + m^2 + \vec \alpha^2+ \vec \delta_2^2 \pm 2
\sqrt{D_1(\vec p)}, \label{dispersion1.1}
\end{equation}
where 
\beq
D_1(\vec p) = (\vec \alpha \times \vec p - m \vec \delta_2)^2 + (\vec \delta_2 \cdot \vec p)^2 + 
(\vec \alpha \cdot \vec \delta_2)^2,\label{dispersion1.2}
\eeq
is a non-negative quantity.

When $\vec {\delta_1} - \vec {\delta_2}=0$ and $\vec{\alpha}=0$ the
associated dispersion relation becomes
\begin{equation}
p_0^2 =   \vec p^2 + m^2 + \alpha_0^2 + \vec \delta_2^2 \pm 2
\sqrt{D_2(\vec p)}, 
\label{dispersion2.11}
\end{equation}
where
\begin{equation}
D_2(\vec p) = (\vec \delta_2 \times \vec p)^2 + (\alpha_0 \vec p  -  m
\vec \delta_2)^2 , \label{dispersion2.2}
\end{equation}
and $D_2(\vec p) \ge 0$ as in the first case.

Note that in both of these special cases, the dispersion relation is
symmetric under $p_0 \rightarrow - p_0$ indicating that positive and
(reinterpreted) negative energy states are degenerate.  In addition,
for a fixed value of $p_0$, the set of solutions for $\vec p$ forms a
deformed sphere with two sheets where the radius as a function of
angle  is determined by the relevant factor, $D_1(\vec p)$, or
$D_2(\vec p).$  This simple geometric interpretation works well
provided that the Lorentz-violation parameters are small relative to
the momentum and mass involved.  Special degeneracies may arise  
when the Lorentz-violating parameters become comparable to the size of the 
momentum or mass involved.  In what follows, the variety along which
the deformed spheres intersect defines a singular set along which
special care must be taken when performing the associated analysis.

\vglue 0.6cm
{\bf \noindent III. LEGENDRE TRANSFORMATIONS}
\vglue 0.4cm
Recall, the Legendre transform is a natural map from $TM$ to $T^*M$
which maps the Lagrangian, $L,$ of a classical system to its corresponding
Hamiltonian, $H.$  The inverse of the Legendre transformation (when it
is defined) can be computed by solving 
\beq
L= \vec p \cdot \vec v -H.
\eeq
Starting with the modified dispersion relations defined by
(\ref{dispersion2.11}) and (\ref{dispersion2.2}), we consider the $\vec
d$-limit of the SME defined by allowing only $d^{0i} \ne 0.$  The
hamiltonian for the SME $\vec d$-limit can be expressed in the simple form 
\beq
H^2 = m^2 + \vec p^2 (1 \pm |\vec d \cdot  \hat p|)^2 .
\eeq
Note that $\sqrt{H^2 - m^2}$ is a homogeneous function of degree one
in $\vec p$.  This implies that 
\beq
p^i {\partial \over \partial p^i} (H^2 - m^2)^{1/2} = (H^2 - m^2)^{1/2}.
\eeq
Defining $v^i = \partial H /\partial p^i$ as usual yields the simple relation
\beq
LH = -m^2.
\label{laghrel}
\eeq
It is interesting to note that equation (\ref{laghrel}) holds whenever $\sqrt{H^2 - m^2}$
happens to be a homogeneous function of the momentum.
For example, choosing the spatial components $c^{ij}$ as the only nonvanishing
coefficients produces an example that satisfies the above relation.
Elimination of the momentum variables in terms of $\vec v
$ yields a quadratic equation for the quantity
\beq
C_{d\pm} = {H^2  \over H^2 - m^2} = \left[ \sqrt{1 - (\hat {v} \times \vec d)^2} 
\pm |\hat {v} \cdot \vec d| \right]^2 .
\eeq
The classical lagrangian then takes the form
\beq
L_{d\pm} = -m \sqrt{1 - {\vec v^2 \over C_{d\pm}}} .
\label{d0ilag}
\eeq

Another limit that produces a very similar dispersion relation is the case
$g^{0ij} \ne 0$ with $g$ antisymmetric in the last two indices.
Defining $g^{0ij} = \ep^{ijk} g^k$ allows the
hamiltonian to be expressed as
\beq
H = m^2 + \vec p^2 (1 \pm |\vec g \times \hat p|)^2 .
\eeq
Calculations similar to those used to establish the $\vec d$-limit of
the SME yield 
\beq
C_{g\pm} = {H^2  \over H^2 - m^2} = \left[ \sqrt{1 - (\hat v \cdot \vec g)^2} 
\pm |\hat v \times \vec g| \right]^2 ,
\eeq
with corresponding classical lagrangian
\beq
L_{g\pm} = -m \sqrt{1 - {\vec v^2 \over C_{g\pm}}}.
\eeq
Note that this lagrangian is in some sense dual to the previous case as
the roles of dot and cross products are simply interchanged.

\vglue 0.6cm
{\bf \noindent IV. FINSLER SPACES}
\vglue 0.4cm

Conversion to Euclidean space converts the lagrangians to homogeneous,
positive definite functions that can be used to define new Finsler geometries.
One way to implement the transition is to use the replacement
\beq
v^j \rightarrow {u^j \over u^0},
\eeq
for the three spatial velocity components, set the mass to unity, and
implement an appropriate Wick rotation to 
yield the parametrization invariant Finsler structure
\beq
F =  \sqrt{(u^0)^2 + {\vec u^2 \over C_\pm(\vec u)}},
\eeq
where $C_\pm(\vec u)$ is one of the $C$-functions mentioned in the 
previous section.
For example, the $\vec g$-term yields
\beq
C_\pm(\vec u) = \left[ \sqrt{1 - (\hat u \cdot \vec g)^2} \pm |\hat u \times \vec g| \right ]^2 .
\eeq
Note that this function does not vanish provided $\vec g^2 \ne 1$.
Restricting $\vec g^2 < 1$ (or $\vec g^2 >1$) suffices to maintain the positive-definite nature of $F$
for all values of non-zero velocity generating a globally positive definite structure.

The Finsler metric is computed using the formula
\beq
g_{ij} = {1 \over 2}{ \partial^2 F^2 \over \partial u^i \partial u^j}.
\eeq
The zero-components of $g$ are $g_{00} = 1$ and $g_{0i} = 0$,
while the spatial components take the form
\beq
g_{ij} = {1 \over  C_\pm D}\left[ (D \mp (\hat u \cdot \vec g)^2)\de_{ij} \pm g_i g_j \right]
\pm {(\hat u \cdot \vec g)^2 \over D^3}g^\perp_i g^\perp_j ,
\eeq
where $D = |\hat u \times \vec g| \sqrt{1 - (\hat u \cdot \vec g)^2}$ and 
$g^\perp_i = g_i - (\hat u \cdot \vec g )\hat u_i$ is the component of $\vec g$ 
perpendicular to $\vec u$.
Note that the metric is homogeneous of degree zero in $\vec u$.
Generalization of this case to higher numbers of spatial dimensions is straightforward
as the cross product magnitude may be expressed in terms of dot products using
\beq
(\vec u \times \vec g)^2 = \vec u^2 \vec g^2 - (\vec u \cdot \vec g)^2 .
\eeq
The metric for either sheet (plus or minus solutions) exhibits a singularity along lines $\hat u$
parallel to $\vec r$ due to the dependence of $D$ in the denominator.  This is due to the fact that 
$F$ has a cusp at these lines and neither sheet is separately differentiable there.  These are also precisely the points where the two sheets are degenerate.  One simple approach is to delete these lines from the manifold and define the Finsler structure on $TM$ \textbackslash $S$, where $S$ is the set of these degenerate points.  

The Cartan torsion, defined by $C_{ijk} = \half \partial g_{ij}/\partial u^k$, takes the form
\beq
C_{ijk} = \pm {\hat u \cdot \vec g \over 2 |\vec u| D^3}\left[
{3 \over D^2}(\vec g^2 - (\hat u \cdot \vec g)^4)g^\perp_i g^\perp_j g^\perp_k - (\hat u \cdot \vec g)^2w_{ijk}
\right],
\eeq
where $w_{ijk} = \sum_{(ijk)} (\delta_{ij} g^\perp_k - \hat u_i \hat u_j g^\perp_k) $, and $(ijk)$ indicates a sum over cyclic permutations of the indices.  The mean Cartan torsion $I_k = g^{ij} C_{ijk}$ can be found by contracting the Cartan torsion with the inverse metric
\beq
g^{ij} = C_\pm D \left[ {1 \over D \mp (\hat u \cdot \vec g)^2} 
(\de^{ij} - \hat u^i \hat u^j - \hat g_\perp^i \hat g_\perp^j) + {D \over (D\pm \vec g_\perp^2) \vec g_\perp^2} U^{ij}
\right] ,
\eeq
where 
\beq
U^{ij} = D \hat g_\perp^i \hat g_\perp^j \mp \hat u \cdot \vec g(g_\perp^i \hat u^j+g_\perp^j \hat u^i) 
+ \left( D \mp (\hat u \cdot \vec g)^2 \pm (1 + {(\hat u \cdot \vec g)^2 C_\pm \over D^2})\vec g_\perp^2\right)
\hat u^i \hat u^j .
\eeq
The Matsumoto torsion $M_{jkl} = C_{jkl} - {1 \over (n+1)}\sum_{(jkl)}I_j h_{kl}$ can then be constructed using the above quantities together with an
angular metric defined by $h_{jk} = g_{jk} - F_{u_j}F_{u_k}$, where
the subscripts again indicate differentiation.
The general expression is not particularly illuminating, however, it can be
shown that it does not generally vanish
which proves (via the Matsumoto-Hojo theorem \cite{mathoj}) that this space is not isomorphic to
a Randers space.

Taking $\vec g$ as a constant background field leads to zero curvature and Christoffel symbols and not much of interest as far as the geometry goes.  The above computation can be generalized
in two natural ways to generate a more interesting geometry.  First, it is possible to replace the 
constant background field by $\vec g(x)$, a vector field depending on the location in $M$.  The
second possibility is to generalize the inner product to include a general Riemannian metric on 
the spatial part of the manifold.  Either of these generalizations will produce non-trivial Christoffel symbols and curvature.  These generalizations are more naturally handled using the covariant approach discussed in the next section.

\vglue 0.6cm
{\bf \noindent V. COVARIANT APPROACH}
\vglue 0.4cm

A covariant expression for the lagrangian can be generated if one starts with 
a covariant set of nonvanishing lorentz-violating coefficients.  
To obtain a solvable case, it is useful to start with a covariant form for $d^\mn$ with the special restriction of antisymmetry
in the two indices.  The dispersion relation can then be put into the form
\beq
(p^2 - m^2 - B)^2 - 4 m^2 B = 0,
\eeq
where $B = d_{\mu \alpha}d^{\al}_{~\nu}  p^\mu p^\nu \equiv (d^2)_\mn p^\mu p^\nu$ depends on the (symmetric) square of the background
tensor.
The derivatives with respect to $p^i$ can be taken implicitly leading to a vector equation
relating the momentum and the velocity.  This equation can be converted into three
different covariant equations by taking the dot product with $p^\mu$, $u^\mu$, $(d^2 p)^\mu$ and
$(d^2 u)^\mu$ and taking appropriate linear combinations of the resulting equations.
The algebra is a bit involved, but in the end it produces an fourth-order polynomial for the 
square of the lagrangian when appropriate zero sets are chosen.  
Borrowing an observation of \cite{aknr} regarding antisymmetric tensors,
\beq
(d^4)_\mn = Y^2 \eta_\mn - 2 X(d^2)_\mn ,
\eeq
is used to reduce higher powers of the antisymmetric tensor.
In this expression, $X = d_\mn d^\mn /4$, and $Y=d_\mn \tilde d^\mn /4$, with $\tilde d$ denoting 
the conventional dual of $d$.
Closed form solutions for the above procedure
exist, but their structure is very complicated and not particularly illuminating, therefore, only a special case is considered here.
  
The polynomial for the lagrangian factors when the covariant quantity $Y=d_\mn \tilde d^\mn /4$
vanishes, where $\tilde d^\mn$ is the standard dual.  
The appropriate zero set then reduces to a
quadratic equation in $L^2$ with particle lagrangian solutions
\beq
L_d = - {m \over 1 - 2 X} \left[ \sqrt{u^2(1 - 2 X) + D} \pm \sqrt{D} \right] ,
\label{covlag}
\eeq
where $D =( d^2)_\mn u^\mu u^\nu$ and $X = d_\mn d^\mn /4$ are covariant quantities
constructed from the background tensor and velocity four-vector.
Antiparticle solutions also exist with opposite sign for the mass.
Note that $D \ge 0$ must hold for the lagrangian to be real.  For example, this is the case
if the time-components of $d^\mn$ vanish and $d^{ij} = \ep^{ijk} d^k$.  In this case a simple computation yields $D =  (\vec u \times \vec d)^2$ and $X = \vec d^2 /2$.
Note that this does not in fact yield the same lagrangian as equation (\ref{d0ilag})
since in that case $d^{i0}=0$ was imposed rather than antisymmetry.
One advantage of the expression (\ref{covlag}) is that it may be generalized to 
an arbitrary background spacetime metric by simply modifying the inner product that
is inherent in the symbols with $\eta_\mn \rightarrow r_\mn(x)$.  The anti-symmetric two-form
can also be promoted to a position-dependent form $d_\mn(x)$ in a natural way.

The momentum can be computed by differentiating the lagrangian yielding
\beq
p_\mu = {m \over \sqrt{u^2(1 - 2 X) + D}}\left( u_\mu 
\mp {L \over m \sqrt{D}} (d^2)_{\mu \al} u^\al \right) .
\eeq
Note that the second term has a finite but undetermined limit as $D \rightarrow 0$.
If this happens in some physical regime there can be interesting effects such as
sudden jumps in the velocity as the momentum 
crosses through certain threshold conditions when a particular sign is chosen
for $F$.  This is not surprising since there are cusp points in momentum space
along these directions when spin is neglected.

A Finsler structure can be defined by Wick rotation as before yielding
the "antisymmetric d-space" finsler structure
\beq
F_d = {1 \over 1 - 2 X} \left( \sqrt{u^2(1 - 2X) + D} \pm \sqrt{D}\right) .
\eeq
It is interesting to note that this is not quite of simple bipartite form 
$F = \sqrt{u^2} \pm \sqrt{u^j s_{jk}u^k}$,
as defined in \cite{kosfins},
but a slightly more general form allowing an additional perturbation in the first term.
This new effect is due to the additional momentum-dependence of the couplings involved.
The metric can also be computed as
\beq
g_\mn = {F \over \sqrt{D} \sqrt{u^2(1 - 2 X) + D}}\left( \sqrt{D} r_\mn \pm F (d^2)_\mn \right)
\pm {u^2 \over \left(D(u^2(1 - 2 X) + D)\right)^{3 \over 2}} X_\mn,
\eeq
where $X_\mn = D \left[ (d^2)_{\mu \al} u^\al u_\nu + (d^2)_{\nu \al} u^\al u_\mu \right] 
- u^2 (d^2)_{\mu\al}  (d^2)_{\nu \be} u^\al u^\be$.

\vglue 0.6cm
{\bf \noindent VI. SUMMARY}
\vglue 0.4cm

Various limits of the SME have been constructed that lead to relatively simple Finsler
structures.  In this paper, we have presented new special cases involving
momentum-dependent couplings that yield additional Finsler functionals with
some interesting properties.  
Various non-covariant choices for the background fields yield solvable Legendre
transformations, mainly due to the relatively simple dependence of the hamiltonian on the
momentum.  
For these cases, the Legendre transformation may be explicitly constructed.
The solvable case of antisymmetric $d^{ij}$ suggested attempting a covariant implicit solution
for this case which turned out to be successful.  This example should be easily generalized
to arbitrary riemanian metrics and more general antisymmetric two-forms that depend
on manifold location.  
The full theory of connection coefficients and curvature calculations can then be 
applied to this new family of examples.

Momentum-dependent couplings involving the antisymmetric $d^\mn$-term introduces 
a new generalized bipartite Finsler structure of the form
$F = \sqrt{u^2+ \delta_1} \pm \sqrt{\delta_2}$,
in which both the first and second square roots are perturbed.  
All previously solved simple cases involving momentum
independent terms have yielded $\delta_1=0$, while all previously solved
momentum-dependent terms have yielded $\delta_2=0$, so
this is in fact a new case worthy of future consideration.
It is also likely that some clever choice of $g^{\la\mu\nu}$ may also lead to 
a solvable structure, another case for future work.

\vglue 0.6cm
{\bf \noindent ACKNOWLEDGMENTS}
Support of New College of Florida's faculty
development funds contributed to the successful completion of
this project.

\vglue 0.6cm
{\bf\noindent REFERENCES}
\vglue 0.4cm


\begin{thebibliography}{xx}


\bibitem{ck} 
D.\ Colladay and V.A.\ Kosteleck\'y,
Phys.\ Rev.\ D {\bf 55}, 6760 (1997);
Phys.\ Rev.\ D {\bf 58}, 116002 (1998).

\bibitem{kle} 
V.A.\ Kosteleck\'y and R.\ Lehnert,
Phys.\ Rev.\ D {\bf 63}, 065008 (2001).

\bibitem{klp1}
V.A.\ Kosteleck\'y, C.\ Lane, and A.\ Pickering,
Phys.\ Rev.\ D {\bf 65}, 056006 (2002).

\bibitem{klp2}
V.A.\ Kosteleck\'y and A.\ Pickering,
Phys.\ Rev.\ Lett.\ {\bf 91}, 031801 (2003).

\bibitem{cm1}
D.\ Colladay and P.\ McDonald,
Phys.\ Rev.\ D {\bf 75}, 105002 (2007);
Phys.\ Rev.\ D {\bf 77}, 085006 (2008);
Phys.\ Rev.\ D {\bf 79}, 125019 (2009).

\bibitem{kosgrav}
V.\ A.\ Kosteleck\'y,
Phys.\ Rev.\ D {\bf 69} 105009 (2004).

\bibitem{kpott}
A.\ Kosteleck\'y and R.\ Potting,
Phys\ Rev.\ D {\bf 79}, 065018 (2009).

\bibitem{kmewes}
A.\ Kosteleck\'y and M.\ Mewes,
Phys.\ Rev.\ Lett.\ {\bf 99}, 011601 (2007).

\bibitem{cmm}
D.\ Colladay, P.\ McDonald, and D.\ Mullins,
J.\ Phys.\ A: Math.\ Theor.\  {\bf 43} 275202 (2010).

\bibitem{neil}
V.\ A.\ Kosteleck\'y and N.\ Russell
arXiv:0801.0287 (2011).

\bibitem{cpt98}
For a summary of recent theoretical models and
experimental tests
see, for example,
{\it CPT and Lorentz Symmetry IV}, V.A.\ Kosteleck\'y, ed.,
World Scientific, Singapore, 2008;
{\it CPT and Lorentz Symmetry V}, V.A.\ Kosteleck\'y, ed., 
World Scientific, Singapore, 2011.

\bibitem{randers}
G.\ Randers,
Phys.\ Rev.\ {\bf 59}, 195 (1941).

\bibitem{kosfins}
V.\ A.\ Kosteleck\'y,
Phys.\ Lett.\ B {\bf 701}, 137 (2011).

\bibitem{aknr}
V.\ A.\ Kosteleck\'y and N.\ Russell,
Phys.\ Lett.\ B {\bf 693}, 443 (2010).

\bibitem{opera}
T.\ Adam, et. al, arXiv: 1109.4897v1 (2011).

\bibitem{colglash}
A.G.\ Cohen and S. L.\ Glashow,
Phys.\ Rev.\ Lett.\ {\bf 97} 021601 (2006).

\bibitem{brett}
B.\ Altschul,
Phys.\ Rev.\ D {\bf 84} 091902 (2011).

\bibitem{kosneut}
V.\ A.\ Kosteleck\'y and M.\ Mewes,
arXiv:1112.6395 (2011).

\bibitem{clw}
Z.\ Chang, X.\ Li, and S.\ Wang,
arXiv: 1110.6673 (2011); arXiv:1201.1368 (2012).

\bibitem{romero}
J.\ M.\ Romero, O.\ S.\ Santos, and J.\ D.\ Vergara,
arXiv:1106.3529 (2011).


\bibitem{BCS}
D.\ D.\ Bao, S.\ Chen, and Z.\ Shen,
{\it An Introduction to Riemann-Finsler Geometry},
Springer, New York, (2000).

\bibitem{ZS}
Z.\ Shen,
{\it Differential Geometry of Spray and Finsler Spaces},
Kluwer Academic Publishers,
Dordrecht, (2001).

\bibitem{Bo1}
G.\ Y.\ Bogoslovsky,
Phys.\ Lett.\ A {\bf 350}, 5 (2006).

\bibitem{Bo2}
G.\ Y.\ Bogoslovsky,
SIGMA {\bf 1} 017 (2005).

\bibitem{mathoj}
M.\ Matsumoto, Tensor, NS {\bf 24}, 29 (1972);
M.\ Matsumoto and S. Hojo, Tensor, NS {\bf 32}, 225 (1978).



\end{thebibliography}
\end{document}